\documentclass[%
 reprint,
 amsmath,amssymb,
 aps,
]{revtex4-2}

\usepackage{graphicx}
\usepackage{dcolumn}
\usepackage{bm}
\usepackage{xcolor}
\usepackage{amsmath,amssymb}
\usepackage{amsfonts}
\usepackage{soul}

\newcommand\matiassout{\bgroup\markoverwith{\textcolor{teal}{\rule[.5ex]{2pt}{0.8pt}}}\ULon}

\usepackage[english]{babel}

\begin{document}
\preprint{APS/123-QED}

\title{On Geometric Cosmology}





\author{Gustavo Arciniega}
 \email{gustavo.arciniega@ciencias.unam.mx}
\affiliation{Departamento de F\'{\i}sica, Facultad de Ciencias, Universidad Nacional Aut\'onoma de M\'exico, Apartado Postal 50-542, 04510, CDMX, M\'exico}

\author{Luisa G. Jaime}
 \email{luisa@ciencias.unam.mx}
\affiliation{Departamento de F\'{\i}sica, Facultad de Ciencias, Universidad Nacional Aut\'onoma de M\'exico, Apartado Postal 50-542, 04510, CDMX, M\'exico}

\author{Susana J. Landau}
\email{slandau@df.uba.ar}
\affiliation{Departamento de   Física, Facultad de Ciencias Exactas y Naturales, Universidad de Buenos Aires, \\
Av. Intendente Cantilo S/N, 1428,  Ciudad Autónoma de Buenos Aires, Argentina }
\affiliation{IFIBA - CONICET - UBA \\
Avenida Intendente Cantilo S/N, 1428, Ciudad Autónoma de Buenos Aires, Argentina}
\author{Matías Leizerovich}
\email{mleize@df.uba.ar}
\affiliation{Departamento de Física, Facultad de Ciencias Exactas y Naturales, Universidad de Buenos Aires \\
Av. Intendente Cantilo S/N 1428  Ciudad Autónoma de Buenos Aires, Argentina }
\affiliation{Consejo Nacional de Investigaciones Científicas y Técnicas (CONICET), Godoy Cruz 2290, 1425, Ciudad Autónoma de Buenos Aires, Argentina}

\date{\today}

\begin{abstract}
We present a modification to General Relativity by making a redefinition of the coupling constant in front of the Ricci curvature scalar along with the Generalized Quasi-topological Gravity theories added to the action, that we named Geometric Cosmology. We give four different exponential convergent models for this class of theories belonging to three different gravities of the Geometric Cosmology theories. 
\end{abstract}

\keywords{Geometric inflation, Geometric acceleration, Modified gravity, Cosmology}


\maketitle


\section{Introduction}

Recently the DESI collaboration has published a new data release \cite{DESI:2025fxa} where they report that: ``Unless there is an unknown systematic error associated with one or more datasets, it is clear that $\Lambda$CDM is being challenged by the combination of DESI BAO with other measurements and that dynamical dark energy offers a possible solution'' \cite{DESI:2025results}. Even though there are some problems that better observations are alleviating, see for example the $\sigma_8$-tension \cite{Wright:2025sigma8}, there are some other observational results that have a certain disagreement with the current cosmological standard model $\Lambda$CDM \cite{Perivolaropoulos:2021Chalenges}, being the $H_0$-tension the most stringent \cite{Kamionkowski:2022pkx}, but there are also theoretical issues like the lack of understanding of the nature of the inflaton and dark energy \cite{Oks:DM-DE} that make us consider the possibility of a modified gravity theory as a better alternative to model the evolution of the universe. In that sense, many alternatives have been proposed by the community (for a review of modified gravity in cosmology see \cite{CANTATA:2021asi} and references in it). 

In recent years, the gravitational theories known as Generalized Quasi-topological Gravities (GQTG) \cite{Bueno:2019GQTG, Moreno:2023GQTG} have appeared as an alternative theory to General Relativity (GR). The extension to include cosmology is known as Cosmological GQTG (CGQTG)\cite{Moreno:2023CGQTG} and has been proved, at least at the theoretical level, to be consistent with the criteria for a new cosmological theory of gravity (\cite{Arciniega:GI, Arciniega:ensayo, Arciniega:gabriella, Quiros:2020uhrCECG, Quiros:2020eim, Jaime:viability, Jaime:AUnified}).

There is a subtle generalization to the CGQTG theories that has not been taken before in the literature, but that has some curious properties like taking the Einstein constant $\kappa$ as a free parameter that, even, can be zero. The former forces us to consider the higher order curvature densities of the CQGTG theory as the sole source of gravity. This is why we will explore the theoretical framework of this class of theories compared with the CGQTG known models, and, in a second paper, we will perform the statistical cosmological analysis \cite{Matias}.

In section \ref{sec:generalAction}
 we present the most general action constructed as a linear combination of Lagrangian densities $\mathcal{R}^{(n)}$ which are formed by the contraction of $n$ curvature tensors. We make a short review of the theories belonging to that kind of action, in particular, the one coined as Generalized Quasi-topological Gravity, and we define what a Geometric Cosmology is. In section \ref{sec:cosmologyGC} we focus on the field equations of motion for a Friedmann-Lema$\hat{\i}$tre-Robertson-Walker metric for the proper Generalized Quasi-topological Gravity theories and focus on the exponential convergent function that is named GILA model. In section \ref{sec:beta-deformation} we make the same as the previous section but now for the theory $\beta$-deformation. In section \ref{sec:pureGC} we present two exponential convergent functions that present a contribution similar to the GILA and $\beta$-deformation exponential models, but with no contribution of the linear Ricci curvature scalar of General Relativity. At the end, we give our conclusions and perspectives. 

\section{The general action}
\label{sec:generalAction}

The most general action constructed by adding higher-order Lagrangian densities with contractions of curvature tensors with no covariant derivatives is:

\begin{equation}\label{actionGeneral}
    S=\frac{1}{2\kappa}\int d^4x  \sqrt{-g}\left(\alpha_1 R+\sum^{\infty}_{n=2}\alpha_{(n,i)}\mathcal{R}^{(n)}\right),
\end{equation}

\noindent where $\kappa=(8\pi G)/c^4$, $\alpha_{(n,i)}$ are dimensional coupling constants of the theory, $n$ is the number of curvature tensors used to construct the Lagrangian density, and $i$ is the number of different Lagrangian densities of order $n$. The $\mathcal{R}^{n}$ are $n$-curvature tensor contractions forming a Lagrangian density for each $n$.

For example, for $n=2$, $\alpha_{(2,i)}\mathcal{R}^{(2)}=\alpha_{(2,1)}R^2+\alpha_{(2,2)}R_{ab}R^{ab}+\alpha_{(2,3)}R_{abcd}R^{abcd}$, where the $i$ are the three curvature contraction tensors that can be constructed using two tensors of curvature. For $n=3$ we have eight different Lagrangian densities: 
\begin{eqnarray}\nonumber
\alpha_{(3,i)}\mathcal{R}^{(3)}&=&\alpha_{(3,1)}R_{a}{}^{c}{}_{b}{}^{d}R_{c}{}^{e}{}_{d}{}^{f}R_{e}{}^{a}{}_{f}{}^{b}\\ \nonumber
&+&\alpha_{(3,2)}R_{ab}{}^{cd}R_{cd}{}^{ef}R_{ef}{}^{ab}\\ \nonumber
&+&\alpha_{(3,3)}R_{abcd}R^{abc}{}_{e}R^{de}+\alpha_{(3,4)}RR_{abcd}R^{abcd}\\ \nonumber
&+&\alpha_{(3,5)}R_{abcd}R^{ac}R^{bd}+\alpha_{(3,6)}R_a{}^bR_b{}^cR_c{}^a\\
&+&\alpha_{(3,7)}RR_{ab}R^{ab}+\alpha_{(3,8)}R^3. 
\end{eqnarray}

For $n=4$ there are 26, and so on (see \cite{Fulling:1992vm}).

\subsection{Recovering GR}

When $\alpha_1\neq 0$, and $\alpha_{n}=0$ for all $n\geq 2$, we recover General Relativity (GR) for $\alpha_1=1/(2\kappa)$. In any other case, for a general $\alpha_1\neq 0$, we can call it a deformation of GR, where $\kappa\rightarrow \kappa_{\rm eff}$, meaning that we have a Newton effective gravitational constant $G_{\rm eff}$. In this case, action (\ref{actionGeneral}) is reduced to:

\begin{equation}\label{actionGRdeformed}
    S=\int d^4x  \sqrt{-g}\left(\frac{1}{2\kappa} R\right)|_{\kappa_{\rm eff}=\kappa}.
\end{equation}

\subsection{Gravitational theories with higher-order Lagrangian densities}

The general action (\ref{actionGeneral}) has infinite degrees of freedom. In this section, we are going to define some of the most known gravitational theories that came from action as given in (\ref{actionGeneral}).

\subsubsection{$f(R)=R+\alpha R^{n}$ modified gravitational theory}

From action (\ref{actionGeneral}), if we only take the Ricci scalar Lagrangian density, we have a specific $f(R)$ modified gravity theory, read \cite{Starobinsky:2007fR}:

\begin{equation}
    S=\frac{1}{2\kappa}\int d^4x \sqrt{-g}(R+\alpha R^n).
\end{equation}

Even if $f(R)$ theories are not of interest to this article and do not belong to the class of theories called Geometric Cosmology, it does pertain to the class of gravity arising from action (\ref{actionGeneral}). We wanted to mention them to emphasize that the general action we have taken at the beginning of this manuscript contains a complete family of different theories; most of them with equations of motion of fourth order. In particular, for the $f(R)$ modified gravity, the field equations read as \cite{Jaime:2017fR}:

\begin{equation}
    f_RR_{ab}-\frac{1}{2}fg_{ab}-\Big(\nabla_a\nabla_b -g_{ab}\square \Big)f_R=\kappa T_{ab},
\end{equation}

\noindent where $f_R=\partial_R f$, $\square=g^{ab}\nabla_a\nabla_b$ is the covariant D'Alembertian, and $T_{ab}$ is the stress-energy tensor.

\subsection{Generalized Quasi-topological Gravitiy}

Following previous authors \cite{Bueno:2019GQTG, Bueno:2022resGQTGwhole, Moreno:2023GQTG}, Generalized Quasi-topological Gravities (GQTGs) are gravitational theories whose action can be written as a finite or infinite tower of Lagrangian densities of curvature invariants, as has been done in equation (\ref{actionGeneral}), and fulfill the following properties:

\begin{itemize}

\item The linearized field equations around any maximally symmetric spacetime only propagates the transverse and traceless graviton in a vacuum, up to a redefinition of the Newton constant \cite{Bueno:2016ECG, Bueno:2016Aspects}. 

\item The equations of motion for a static spherically symmetric black hole, 

\begin{equation}\label{sss-metric}
    ds^2=-f(r)dt^2+\frac{dr^2}{f(r)}+r^2d\theta^2+r^2\sin^2\theta d\phi^2,
\end{equation}

\noindent are at most of second order \cite{Bueno:2016ECGBH, Bueno:2017BH, Bueno:2017StableBH, Bueno:2017qceStableBH4D}.

\end{itemize}

As we will show next, there are several theories that fulfill the second-order condition for black hole metrics; some of them, such as Lovelock's theories, are well-known in the literature.
 
\subsubsection{Lovelock gravity} 

Lovelock gravity is the most general theory whose equations of motion are of second order for any metric solution of the equations of motion, and it is free of ghosts or any other modes except the massless graviton \cite{Lovelock:1971yv, Lovelock:1972vz}; in particular, for black hole-like solutions, which make them a subset of the GQTGs.

The Lovelock Lagrangian density can be written as:

\begin{equation}
    \mathcal{L}=\sqrt{-g}\sum_{n=0}^{t}\alpha_n \mathcal{R}^{(n)},
\end{equation}

\noindent where

\begin{equation}
    \mathcal{R}^{(n)}=\frac{1}{2^n}\delta^{a_1 b_1\ldots a_n b_n}_{c_1 d_1\ldots c_n d_n}\prod_{r=1}^{n}R^{c_r d_r}{}_{a_r b_r},
\end{equation}

\noindent and where we used the generalized Kronecker  $\delta$-function, defined as the antisymmetric product of deltas:

\begin{equation}
     \delta^{a_1 b_1\ldots a_n b_n}_{c_1 d_1\ldots\ c_n d_n}=\frac{1}{n!}\delta_{[ c_1}^{a_1}\delta_{d_1}^{b_1}\ldots \delta_{ c_n}^{a_n}\delta_{d_n]}^{b_n}. 
\end{equation}

With the above prescription, we can compute the first non-trivial Lovelock Lagrangian densities:

\begin{equation}
    \mathcal{R}^{(0)}=\Lambda,
\end{equation}

\noindent where $\Lambda$ is constant, and it is related to the cosmological constant.

\begin{equation}
    \mathcal{R}^{(1)}=R,
\end{equation}

\begin{equation}\label{gauss-bonnet}
    \mathcal{R}^{(2)}=R^2-4R_{ab}R^{ab}+R_{abcd}R^{abcd},
\end{equation}

\noindent and

\begin{eqnarray}\nonumber
    \mathcal{R}^{(3)}& = & 2{R_{ab}}^{cd} {R_{cd}}^{ef} {R_{ef}}^{ab}-8{{{R_a}^c}_b}^d {{{R_c}^e}_d}^f {{{R_e}^a}_f}^b \\ 
&&-24R_{abcd} {R^{abc}}_e R^{de}+3R_{abcd} R^{abcd} R\\ \nonumber
&&+24R_{abcd} R^{ac} R^{bd}+16{R_a}^b {R_b}^c {R_c}^a\\ \nonumber
&&-12 R_{ab} R^{ab} R+R^3 ~.
\end{eqnarray}

The Lovelock Lagrangian densities are related to the Euler density $\chi_{(2n)}$ which is a topological term in space-times of dimension $2n$. The $n=2$ is the Gauss-Bonnet Lagrangian density, which is a topological term for dimension four, meaning that this Lagrangian does not contribute to the field equations of motion. For $n>2$ and dimension four, the $\chi_{2n}$ Lovelock densities are zero. This is why Lovelock theories are interesting only for higher-dimensional theories like string theory \cite{Garraffo:2008hu, Camanho:2011rj}.

\subsubsection{Quasi-topological Gravity}

Quasi-topological Gravity theories are a subclass of the GQTGs. In these theories, the field equation of motion for a spherically symmetric ansatz $f(r)$, as was given in (\ref{sss-metric}), turns out to be algebraic, i.e. it does not contain derivatives of $f(r)$ \cite{Oliva:2010QTG, Myers:2010BHQTG, Dehghani:2013QTG, Cisterna:2017QTG}. The name came from the fact that some Lagrangian densities behave like a topological term for some particular ansatz, but not for any, as was first found by Myers and Robinson \cite{Myers:2010BHQTG}.

As is written in \cite{Moreno:2023GQTG}, the first three Quasi-topological densities $\mathcal{Z}_{(n)}$ can be written as:

\begin{eqnarray}
    \mathcal{Z}_{(1)}&=&R,\\
    \mathcal{Z}_{(2)}&=&\frac{D(D-1)}{(D-2)(D-3)}\chi_4,\\ \nonumber
    \mathcal{Z}_{(3)}&=&\frac{4(D-1)^2D^2(2D-3)}{(D-3)(D-2)(D((D-9)D+26)-22)}\\
    && \times\Bigg(\mathcal{Z}_{(3)}^{MR}+\frac{1}{8}\chi_6\Bigg),
\end{eqnarray}

\noindent where $\chi_4$ is the Gauss-Bonnet density (\ref{gauss-bonnet}), $\chi_6$ is the cubic Lovelock density given by

\begin{eqnarray}\label{LagChi6} \nonumber
\mathcal{\chi}_6 & = & 2{R_{ab}}^{cd} {R_{cd}}^{ef} {R_{ef}}^{ab}-8{{{R_a}^c}_b}^d {{{R_c}^e}_d}^f {{{R_e}^a}_f}^b \\ 
&&-24R_{abcd} {R^{abc}}_e R^{de}+3R_{abcd} R^{abcd} R\\ \nonumber
&&+24R_{abcd} R^{ac} R^{bd}+16{R_a}^b {R_b}^c {R_c}^a\\ \nonumber
&&-12 R_{ab} R^{ab} R+R^3 ~,
\end{eqnarray}

\noindent and $\mathcal{Z}_{(3)}^{MR}$ is the cubic quasi-topological gravity given in \cite{Myers:2010BHQTG}:

\begin{eqnarray}\nonumber
    \mathcal{Z}_{(3)}^{MR}&=&R_{a}{}^{c}{}_{b}{}^{d}R_{c}{}^{e}{}_{d}{}^{f}R_{e}{}^{a}{}_{f}{}^{b}+\frac{1}{(2D-3)(D-4)} \\ \nonumber
    &&\times \Bigg[\frac{3(3D-8)}{8}RR_{abcd}R^{abcd}-\frac{3(3D-4)}{2}R_a{}^bR_b{}^aR\\ \nonumber
    &&-3(D-2)R_{abcd}R^{abc}{}_{e}R^{de}+3DR_{abcd}R^{ac}R^{bd}\\
    &&+6(D-2)R_{a}{}^{b}R_{b}{}^{c}R_{c}{}^{a}+\frac{3D}{8}R^3\Bigg].
\end{eqnarray}

The authors of \cite{Bueno:2019GQTG} and \cite{Moreno:2023GQTG} found a recurrence relation to write any $\mathcal{Z}_{(n)}$ for any order $n$, in case the reader is interested.

Lovelock's theories are a subset of the Quasi-topological Gravities. That means that, when a Lovelock density is not a topological term or a null contribution to the equations of motion, then it belongs to the Quasi-topological family.

Interestingly, as happens with Lovelock gravity, the quasi-topological theories only exist for dimensional spaces $D\geq 5$.

\subsubsection{Proper Generalized Quasi-topological Gravity}

The proper GQTGs are those for which the field equations for SSS contain derivatives of second order in $f(r)$, to distinguish them from the algebraic equations of motion that characterize the quasi-topological theories of the previous subsection.

The proper GQTGs (pGQTGs) are the only ones that have no trivial contributions in $D=4$. In fact, exist for $D\geq4$. The former makes these theories particularly interesting to analyze as a modified gravity candidate.  

To simplify notation and avoid confusion, in the following, we are going to restrict the discussion to the four-dimensional case $D=4$, which is the one we are interested in this manuscript.

The first pGQTG that has ever been formulated was for a cubic theory and was given in \cite{Bueno:2016ECG}. The action (\ref{actionGeneral}) is reduced to:

\begin{equation}\label{actionPablosECG}
    S=\int d^4x\sqrt{-g}\Big[\frac{1}{2\kappa}(R-2\Lambda)+\beta\mathcal{P}\Big]
\end{equation}

\noindent where $\beta$ is a coupling constant and $\mathcal{P}$ density is defined as:

\begin{eqnarray}\nonumber
\mathcal{P} & = & 12 {{{R_a}^c}_b}^d {{{R_c}^e}_d}^f {{{R_e}^a}_f}^b + {R_{ab}}^{cd} {R_{cd}}^{ef} {R_{ef}}^{ab} \\
&& -12 R_{abcd} R^{ac} R^{bd} + 8 {R_a}^b {R_b}^c {R_c}^a ~. \label{LagP}
\end{eqnarray}

The Einstein constant $\kappa$ is modified for the cubic theory by an effective $\kappa_{eff}$ given by:

\begin{equation}\label{kappa-eff}
    \kappa_{eff}=\frac{\kappa}{1-4\beta}.
\end{equation}

The quartic pGQTG was constructed in \cite{Bueno:2016Aspects}, and was shown to be non-unique:

\begin{equation}
    S=\int d^4x\sqrt{-g}\Big[\frac{1}{2\kappa}(R-2\Lambda)+\gamma_i\mathcal{Q}_i\Big],
\end{equation}

\noindent where $i=1,2,3$, $\gamma_i$ is the $i$th coupling constant and the three densities $\mathcal{Q}_i$ are given by\footnote{The reader can find another expression for the three linear combinations in reference \cite{Cano:2020oaa}.}:

\begin{eqnarray}\nonumber 
    \mathcal{Q}_1&=&3R^{abcd}R_{ab}{}^{ef}R_{ef}{}^{jk}R_{cdjk}-15(R_{abcd}R^{abcd})^2\\ \nonumber
   && -8RR_{a}{}^{c}{}_{b}{}^{d}R_{c}{}^{e}{}_{d}{}^{f}R_{e}{}^{a}{}_{f}{}^{b}+144R^{ab}R^{cd}R_{ac}{}^{ef}R_{efbd}\\ \nonumber
   && -96R^{ab}R_{b}{}^{c}R^{efj}{}_{a}R_{efjc}-24RR_{abcd}R^{ac}R^{bd}\\ \label{Q1}
   && +24(R_{ab}R^{ab})^2,\\ \nonumber
   \mathcal{Q}_2&=&3(R_{abcd}R^{abcd})^2+16 RR_{a}{}^{c}{}_{b}{}^{d}R_{c}{}^{e}{}_{d}{}^{f}R_{e}{}^{a}{}_{f}{}^{b}\\ \label{Q2}
   && -6R^{ab}R_{b}{}^{c}R^{efj}{}_{a}R_{efjc}-60RR_{abcd}R^{ac}R^{bd}\\ \nonumber
   &&-6R_{a}{}^{b}R_{b}{}^{c}R_{c}{}^{d}R_{d}{}^{a}+51(R_{ab}R^{ab})^2+6RR_{a}{}^{b}R_{b}{}^{c}R_{c}{}^{a},\\ \nonumber
   \mathcal{Q}_3&=& R^4+57(R_{abcd}R^{abcd})^2-120 R_{ab}R^{ab}R_{cdef}R^{cdef}\\ \nonumber
   && +6R^2R_{abcd}R^{abcd}-240 RR_{abcd}R^{ac}R^{bd}-144(R_{ab}R^{ab})^2\\ \nonumber
   &&+416 RR_{a}{}^{b}R_{b}{}^{c}R_{c}{}^{a}-24R^2 R_{ab}R^{ab}\\ \label{Q3}
   &&+340 RR_{a}{}^{c}{}_{b}{}^{d}R_{c}{}^{e}{}_{d}{}^{f}R_{e}{}^{a}{}_{f}{}^{b}.
\end{eqnarray}

As happened for the Quasi-topological case, there exist some recurrence relations to write any $\mathcal{R}^{(n)}$ of the pGQTG density at any order $n$ \cite{Bueno:2019GQTG, Moreno:2023GQTG}.

It is worth mentioning that GQTG has $n-1$ inequivalent types of theories for dimension $D\geq5$, one of them is the Quasi-topological theory and the $n-2$ are of the pGQTGs kind \cite{Bueno:2022resGQTGwhole}. Interestingly, for $D=4$ the authors of \cite{Moreno:2023GQTG} proved that there is only one pGQTG, i.e. the $n-2$ different theories that can be constructed at order $n$ give the same field equations of motion for the SSS ansatz (\ref{sss-metric}).

\subsubsection{Cosmological GQTG (Geometric Cosmology)}

Considering now a Friedmann-Lema$\hat{i}$tre-Robertson-Walker metric (FLRW) are:

\begin{equation}\label{FLRW-metric}
    ds^2=-c^2dt^2+a^2(t)\Bigg[\frac{dr^2}{1-kr^2}+r^2d\theta^2+r^2\sin^2\theta \, d\phi^2\Bigg],
\end{equation}

\noindent where $k$ is the spatial curvature, $k=\{-1,0,1\}$, and $a(t)$ the scale factor. 

If the field equations of motion for a FLRW metric of a GQTG are second order, then the theory is said to be a Cosmological GQTG \cite{Moreno:2023CGQTG}. In the following, we are going to refer to a CGQTG as Geometric Cosmology (GC) for short.

The first GC was constructed in \cite{Arciniega:towards} for the cubic Lagrangian density.

The complete cubic action is written as:

\begin{equation}\label{actiontowardsFull}
    S=\int d^4x\sqrt{-g}\Big[\frac{1}{2\kappa}(R-2\Lambda)+ \beta(\mathcal{P}-8\mathcal{C})+\gamma\mathcal{C}'+\frac{1}{2}\chi_{6}\Big],
\end{equation}

\noindent where $\mathcal{P}$ is the same as equation (\ref{LagP}), $\chi_6$ is the Lovelock density given in equation (\ref{LagChi6}), and $\mathcal{C}$ and $\mathcal{C}'$ are defined by the following linear combination of cubic densities:

\begin{eqnarray}
    \mathcal{C} & = & R_{abcd} {R^{abc}}_e R^{de} - \frac{1}{4} R_{abcd} R^{abcd} R \\ \nonumber
&& - 2 R_{abcd} R^{ac} R^{bd} + \frac{1}{2} R_{ab} R^{ab} R ~,\label{LagC} \\ [0.4em] \label{CprimeLag}
\mathcal{C}' & = & 8 {R_a}^b {R_b}^c {R_c}^a - 6 R_{ab} R^{ab} R +R^3 ~. \label{LagCp}
\end{eqnarray}

In a four-dimensional spacetime, the Lagrangian density $\mathcal{C}'$ satisfies  $\mathcal{C}'=4\mathcal{C}$. Taking the former and omitting the density $\chi_6$
that does not contribute to the field equations of motion, action (\ref{actiontowardsFull}) reduces to:

\begin{equation}\label{actiontowards}
    S=\int d^4x\sqrt{-g}\Bigg[\frac{1}{2\kappa}(R-2\Lambda)+\beta(\mathcal{P}-8\mathcal{C})\Bigg].
\end{equation}

Comparing the last action (\ref{actiontowards}) with (\ref{actionPablosECG}), we see that cubic CG is an extension of the pGQTG, however, the $\mathcal{C}$ density is null when it is computed for a SSS ansatz. This cubic CG is the only theory that modifies General Relativity and satisfies the conditions for a pGQTG and cosmology.

In this case, the effective $\kappa_{\rm eff}$ is:

\begin{equation}\label{kappa-eff-towards}
    \kappa_{\rm eff}^{-1}=\kappa^{-1}+48\beta \Lambda.
\end{equation}

The quartic\footnote{For a different expression of the three quartic lagrangian densities see \cite{Cano:2020oaa}} and quintic GC densities are the following \cite{Cisterna:2018tgx}:

\begin{eqnarray}
{\mathcal R}_{(4)} &=& - \frac{1}{192} \bigg[5 R^4 - 60 R^2 R_{ab}R^{ab} + 30 R^2 R_{abcd} R^{abcd} \\ \nonumber
&& - 160 R R_a{}^b{}_c{}^d R_b{}^e{}_d{}^f R_e{}^a{}_f{}^c + 32 R R_{ab}^{cd} R_{cd}^{ef} R_{ef}^{ab} 
\\ \nonumber
&&- 104 R R_{abcd} R^{abc}{}_e R^{de} + 272 (R_{ab}R^{ab})^2  \\ \nonumber
&&- 256 R_{ab}R^{ab} R_{cdef} R^{cdef} + 45 (R_{abcd} R^{abcd})^2 
\\ \nonumber
&&-240 R_{a}{}^{e}{}_{c}{}^{f} R^{abcd} R_{bjdh} R_{e}{}^{j}{}_{f}{}^{h}
\\ && \nonumber	+ 336 R^{ab} R_{a}{}^{c}{}_{b}{}^{d} R_{efhc} R^{efh}{}_{d} + 48 R^{ab} R^{cd} R_{ecfd} R^{e}{}_{a}{}^{f}{}_{b}  \bigg]	
	\\  \nonumber
{\mathcal R}_{(5)} &=& -\frac{1}{5760} \bigg[ 15R^{5}-36R^{3}R_{ab}R^{ab}-224R^{3}R_{abcd} R^{abcd}
\\ \nonumber
&&-336R^{2}R_a{}^b{}_c{}^d R_b{}^e{}_d{}^f R_e{}^a{}_f{}^c
-140R^{2}R_{ab}^{cd} R_{cd}^{ef} R_{ef}^{ab}
\\ \nonumber
&& +528R^{2}R_{abcd} R^{abc}{}_e R^{de}-592R(R_{ab}R^{ab})^{2}
\\ \nonumber
&&+1000RR_{ab}R^{ab}R_{cdef} R^{cdef}+301R(R_{abcd} R^{abcd})^{2}
\\	\nonumber
	&&-912RR_{a}{}^{e}{}_{c}{}^{f} R^{abcd} R_{bjdh} R_{e}{}^{j}{}_{f}{}^{h}
 \\ 
&&-928RR^{ab} R_{a}{}^{c}{}_{b}{}^{d} R_{efhc} R^{efh}{}_{d}
  \\	\nonumber
	&&+1680RR^{ab} R^{cd} R_{ecfd} R^{e}{}_{a}{}^{f}{}_{b} 
 \\ \nonumber
&&+1152R_{ab}R^{ab}R_r{}^s{}_c{}^d R_s{}^e{}_d{}^f R_e{}^r{}_f{}^c
 \\ \nonumber
&&+264R_{ab}R^{ab}R_{rs}^{cd} R_{cd}^{ef} R_{ef}^{rs}+312R_{abcd} R^{abcd}R_{rs}^{cd} R_{cd}^{ef} R_{ef}^{rs}
\\ \nonumber
&&-
64R_{ab}R^{ab}R_{rscd} R^{rsc}{}_e R^{de}
\\ \nonumber
&&-2080R_{abcd} R^{abcd}R_{rscd} R^{rsc}{}_e R^{de}
 \\ \nonumber
 &&+4992R_{ce}{}^{ae} R_{af}{}^{cd} R_{gi}{}^{ef} R_{bj}{}^{gh}R_{dh}{}^{ij}\bigg]	
		 \, .
\end{eqnarray}

For a prescription on the construction of any higher curvature invariant $\mathcal{R}^{(n)}$ see reference \cite{Moreno:2023CGQTG}.

Notice that, in comparison with the quartic pGQTG, the quartic GC is only one linear combination of densities of fourth order.

\section{The cosmology of Geometric Cosmology}\label{sec:cosmologyGC}

The Geometric Cosmology theories are the only theories that (1) are compatible with action (\ref{actionGeneral}), (2) modify Einstein's field equations in dimension $D=4$, and (3) satisfy the condition of second-order equations of motion for a FLRW metric, we are going to focus on this gravitational setup.

\subsubsection{The standard Geometric Cosmology action}

For a GC gravity, the general action (\ref{actionGeneral}) can be written as:

\begin{equation}\label{actionGC}
    S=\frac{1}{2\kappa}\int d^4x  \sqrt{-g}\left( R+\sum^{\infty}_{n=3}\alpha_{n}\mathcal{R}^{(n)}\right).
\end{equation}

There are some differences from the general action that are worth mentioning before continuing. 

First, we have omitted the subindex $i$ in the $\alpha_n$'s. That is because the Lagrangian densities $\mathcal{R}^{(n)}$ are fixed completely by the conditions that are forced to satisfy the GC theories. Thus, there is just one coupling constant $\alpha$ for each order $n$.

Second, the summation starts at $n=3$. We have to remember that the $n=2$ case corresponds to the Gauss-Bonnet topological term in dimension $D=4$. 

Third, we have fixed $\alpha_1=1$ and $\Lambda=0$, i.e., following \cite{Jaime:AUnified}, we are not considering a cosmological constant in the theory.

And fourth, even if we wrote $\kappa$, it is important to keep in mind that it can be shown that the Einstein constant is modified in a specific way to an effective gravitational constant, i.e. $\kappa\rightarrow\kappa_{\rm eff}$, in the same way that was done in (\ref{kappa-eff-towards}). This effective constant will be crucial later in this manuscript, and we will say more when we come back to this point in section \ref{equivalence}.

\subsubsection{Field equations of motion}

We already know that when is performed the variation of the Einstein-Hilbert action containing the linear Ricci scalar with respect to the inverse FLRW metric $g^{ab}$ (from eq. \ref{FLRW-metric}), the 00-component of the field equations is the square of the Hubble parameter, $H^2$, where $H\equiv \dot{a}/a$, and $\dot{(\,)}=d/dt$. Surprisingly, and despite the long and cumbersome expression of the field equations for each $\mathcal{R}^{(n)}$, when these are evaluated for an FLRW ansatz, each density gives an $H^{2n}$ term, where $n$ is the order of the curvature GC density. Then, the modified Friedmann equations for a perfect fluid with energy density $\rho$ and pressure $P$ are \cite{Arciniega:GI, Arciniega:ensayo, Arciniega:gabriella, Jaime:viability, Jaime:AUnified, Cisterna:2024ksz}:

\begin{eqnarray}\label{standard-GC-eom}
    3F(H)&=&\kappa \rho,\\ \label{standard-GC-eom2}
    -\frac{\dot{H}}{H}F'(H)&=&\kappa (\rho+P),
\end{eqnarray}

\noindent where $F'(H)\equiv \partial_H F(H)$, and

\begin{equation}\label{Fdehserie}
    F(H)=H^2+\sum_{n=3}^\infty \alpha_{n}H^{2n}.
\end{equation}

\subsubsection{Convergent GILA exponential models}\label{sec:GILA-exponential}

The series of $F(H)$ can converge to different functions depending on the conditions taken for the coupling constants $\alpha$. In the literature, the exponential convergent functions have been the most explored \cite{Arciniega:GI, Arciniega:ensayo, Arciniega:gabriella, Cisterna:2018tgx, Jaime:viability, Cisterna:2024ksz}. In particular, the authors in \cite{Jaime:AUnified} have split the $F(H)$ series into two infinite series that converge to an exponential function for each one.

The Geometric Inflation (GI) series  is:

\begin{equation}\label{GI-series}
\lambda L^{2(p-1)}H^{2p} \sum_{n=0}^{\infty}\frac{\lambda^{n}L^{2qn}H^{2qn}}{n!}   = \lambda L^{2(p-1)}H^{2p}e^{\lambda(LH)^{2q}},
\end{equation}

\noindent and the Late-time Acceleration (LA) series is:

\begin{multline}\label{LA-series}
\beta \L^{2(r-1)}H^{2r} \sum_{m=0}^{\infty}\frac{(-1)^{m+1}\beta^{m}\L^{2sm}H^{2sm}}{m!}   \\= \beta \L^{2(r-1)}H^{2r}e^{\lambda(\L H)^{2s}}.
\end{multline}

Notice that, for the convergent series GI and LA to belong to the standard Geometric Cosmology, the $p$, $q$, $r$ and $s$ powers have to satisfy:

\begin{equation}\label{condicionesGILA}
  p,q,r,s\in \mathbb{N}, \quad  p,r \geq 3,\quad{\rm and }\quad q,s\geq1,
\end{equation}

\noindent and, taking:

\begin{multline}
    \alpha_n=\Bigg(\lambda L^{2(p-1)}\frac{\beta^nL^{2(q-1)}}{n!}\Bigg)\\
    +\Bigg(\beta\L^{2(r-1)}\frac{(-1)^{n+1}\beta^n\L^{2(s-1)}}{n!}\Bigg)
\end{multline}

\noindent we recover the series expansion for $F(H)$ (\ref{Fdehserie}).

Then, the convergent $F(H)$ for GILA exponential models is

\begin{equation}\label{GILA-convergent}
    F(H)=H^2+\lambda L^{2p-2} H^{2p} e^{\lambda (LH)^{2q}}-\beta \tilde{L}^{2r-2}H^{2r}e^{-\beta (\tilde{L}H)^{2s}},
\end{equation}

\noindent with conditions (\ref{condicionesGILA}).

\section{Effective deformation of GR for Geometric Cosmology}\label{sec:beta-deformation}

Let us consider $\alpha_1$ in action (\ref{actionGeneral}) as a free coefficient, but obeying the conditions for the theory to be a Geometric Cosmology, i.e. second order for equations of motion of a black hole-like and a FLRW metric. Letting $\alpha_1$ be arbitrarily fixed has not been done before in the literature for these kinds of theories. There are two reasons for not considering an arbitrary $\alpha_1$: (1) a nice characteristic of the theory is to have a continuous limit to GR when the coupling coefficients of the modification go to zero, and (2) for a generic $\alpha_i\neq0$ the theory specifies a new $\kappa_{\rm eff}$ which is related by construction with a specific $\alpha_1$. However, to do so enriches the possibilities of modifying GR as a new GC theory. 

\subsubsection{The $\beta$-deformation action}

According to the previous paragraph, let us consider the following action:

\begin{equation}\label{action-beta}
    S=\frac{1}{2\kappa}\int d^4x  \sqrt{-g}\left[(1-\beta) R+\sum^{\infty}_{n=3}\alpha_{n}\mathcal{R}^{(n)}\right],
\end{equation}

\noindent where we have renamed the coupling constant $\alpha_1\rightarrow (1-\beta)$ to facilitate the apparent deviation from the standard GR.


We name the last action as $\beta$-deformation, as we are deforming the Ricci curvature scalar $R$ of GR, and also to avoid confusion with the standard GC case. 

\subsubsection{Field equations of motion}

As the GR modification is on a constant factor, then all the healthy characteristics of GC are the same, including the modified Friedmann equations (\ref{standard-GC-eom}). However, in this case, the $F(H)$ series expansion is written as

\begin{equation}\label{Fbetaserie}
    F(H)=(1-\beta)H^2+\sum_{n=3}^\infty \alpha_{n}H^{2n}.
\end{equation}

\subsubsection{Convergent $\beta$-deformation exponential models}

As has been done in subsection (\ref{sec:GILA-exponential}), we can split the $F(H)$ into two potential infinite series using the same relations as GI (\ref{GI-series}) and LA (\ref{LA-series}), but, in this case, the conditions for $p$, $q$, $r$, and $s$ are:

\begin{equation}\label{condicionesbeta}
   p \geq 3,\quad q\geq1, \quad 
    s\geq2 \quad {\rm and} \quad r=1.
\end{equation}

Then, instead of the convergent function for $F(H)$ in (\ref{GILA-convergent}), we have

\begin{equation}\label{beta-convergent}
    F(H)=H^2+\lambda L^{2p-2} H^{2p} e^{\lambda (LH)^{2q}}-\beta H^2 e^{-\beta (\tilde{L}H)^{2s}}.
\end{equation}

Notice that, despite both formulations being built from the same physical assumptions, it is not possible to obtain (\ref{beta-convergent}) from (\ref{GILA-convergent}), and vice versa.  It is also worth mentioning that the expression for the convergent function $F(H)$ for $\beta$-deformation does not contain a scale factor $\L$ alongside the product of $\beta H^2$ in the third term. 

The fact that the expressions for the two convergent functions $F(H)$ of each theory are similar, makes the $\beta$-deformation theory worth testing as a cosmological scenario.

\subsubsection{Equivalence of $\beta$-deformation with a proper GQTG}\label{equivalence}

It is not difficult to show that the $\beta$-deformation theory pertains to a class of GQTG theories.

In order to do it, let us consider an arbitrary $\beta$-deformation theory with fixed $\beta$ value, i.e. $\kappa_\beta\equiv(1-\beta)^{-1}=8\pi G_\beta/c^4$, an effective gravitational constant. Now, let us consider a proper GQGC that is a maximally symmetric spacetime with spacetime curvature $\Lambda\neq0$, i.e.

\begin{equation}\label{actionLambda}
    S_\Lambda=\frac{1}{2\kappa}\int d^4x  \sqrt{-g}\left( R-2\Lambda+\sum^{\infty}_{n=3}\alpha_{n}\mathcal{R}^{(n)}\right).
\end{equation}

When the pGQGC theory is linearized, the field equations of motion can be expressed in terms of some constants, $a$, $b$, $c$, and $e$, whose values depend on the theory and the $\mathcal{R}^{(n)}$ that is considered. In particular, the $\kappa_{\rm eff}$ of the theory is given by \cite{Bueno:2016Aspects}:

\begin{equation}\label{kappa-eff-kappa}
    \kappa_{\rm eff}=\frac{1}{4e-8\Lambda a},
\end{equation}

\noindent where we are considering here only the case of dimension $D=4$. The computation of $e$ and $a$ is not of particular interest here, only one fact: If $\Lambda a=0$, then $\kappa_{\rm eff}=\kappa$, but that only happens if $\alpha_n=0$ for all $n\geq 3$, i.e. there is no modified gravity, or we are working in Minkowskian plane spacetime $\Lambda=0$. Then, for a pGQTG with $\Lambda\neq 0$ there is always $\kappa_{\rm eff}\neq \kappa$. Also, as $\kappa_{\rm eff}^{-1}$ is proportional to $\Lambda$, then we can always find a $\Lambda$ such that the $\beta$-deformation gravity with no curvature constant $\Lambda$ is equivalent at linear order to a pGQTG gravity theory with a $\Lambda \neq 0$.

\section{Geometric Cosmology with no GR contribution}\label{sec:pureGC}

The $\beta$-deformation of the theory makes us wonder about the implications of considering the case when $\beta=1$ in action (\ref{action-beta}). In such a case, the contribution of General Relativity is omitted, and only the higher-order curvature densities play a gravitational role.

It is important to notice that, for the construction of convergent function $F(H)$ in the $\beta$-deformation theory (\ref{beta-convergent}), it was taken the $\alpha_1=(1-\beta)$ and that same $\beta$ to be equal to the one in the infinite series expansion (\ref{LA-series}). So, we have two cases: (1) when  $\beta=1$ for both, $\alpha_1$ and the infinite series, and (2) when $\beta=1$ ($\alpha_1=0$) but $\beta$ in the infinite series an arbitrary value to be determined.

As at the level of the action both cases are codified in it before the $\alpha_n$ are given, we are going to start with the most general setup of the case $\alpha_1=0$.

\subsubsection{The action and field equations of pure Geometric Cosmology}

Let us consider the $\beta$-deformation action (\ref{action-beta}) with ($\beta=1$):

\begin{equation}\label{action-pureGC}
    S=\frac{1}{2\kappa}\int d^4x  \sqrt{-g}\, \left(\sum^{\infty}_{n=3}\alpha_{n}\mathcal{R}^{(n)}\right),
\end{equation}

\noindent where the curvature densities $\mathcal{R}^{(n)}$ belong to the GC gravity.

The field equations of motion are the same as (\ref{standard-GC-eom}) and (\ref{standard-GC-eom2}), but the $F(H)$ is now without the $H^2$ term:

\begin{equation}\label{FPureGCserie}
    F(H)=\sum_{n=3}^\infty \alpha_{n}H^{2n}.
\end{equation}

Interestingly, if the $\alpha_n$ are selected to be

\begin{multline}\label{fH-pureseries1}
    F(H)=\lambda L^{2(p-1)}H^{2p}\sum_{n=0}^\infty \frac{\lambda^n (L H)^{2qn}}{n!}\\
    -\beta H^2\sum_{n=1}^\infty \frac{(-1)^n\beta^n (\L H)^{2sn}}{n!},
\end{multline}

\noindent with conditions

\begin{equation}
    p\geq3, \quad q\geq 1, \quad {\rm and} \quad s\geq2,
\end{equation}

\noindent then, the convergent function $F(H)$ is

\begin{equation}\label{pure1-convergent}
    F(H)=\beta H^2+\lambda L^{2(p-1)}H^{2p}e^{\lambda(LH)^{2q}}-\beta H^2 e^{-\beta(\L H)^{2s}}.
\end{equation}

And, if the $F(H)$ series expansion is:

\begin{multline}\label{fH-pureseries2}
    F(H)=\lambda L^{2(p-1)}H^{2p}\sum_{n=0}^\infty \frac{\lambda^n (L H)^{2qn}}{n!}\\
    -\beta H^2\sum_{n=1}^\infty \frac{(-1)^n\beta^{n-1} (\L H)^{2sn}}{n!},
\end{multline}

\noindent with conditions

\begin{equation}
    p\geq3, \quad q\geq 1, \quad {\rm and} \quad s\geq2,
\end{equation}

\noindent then

\begin{equation}\label{pure2-convergent}
    F(H)=H^2+\lambda L^{2(p-1)}H^{2p}e^{\lambda(LH)^{2q}}- H^2 e^{-\beta(\L H)^{2s}}.
\end{equation}

The change between the first (\ref{fH-pureseries1}) and (\ref{fH-pureseries2}) is only the exponent of the $\beta$ factor inside the sum: for the first case it is $\beta^{n}$, meanwhile for the second one it is $\beta^{n-1}$.

There are a few remarks to make about these last results. First, it is interesting that, despite we have taken out the $H^2$ contribution in the field equations of motion (\ref{fH-pureseries1}) and (\ref{fH-pureseries2}), the infinite series converges to an $F(H)$ that contains an $H^2$, one with a $\beta$ factor multiplying it (\ref{pure1-convergent}), and the other one with the $\beta$ factor only in the exponent of the last exponential function (\ref{pure2-convergent}). Second, the last convergent function $F(H)$ looks very similar to the $\beta$-deformation gravity (\ref{beta-convergent}) despite what we said previously that there is no possibility of obtaining a $\beta$-deformation theory from a standard GC.

\subsubsection{$\beta=1$ case}

After obtaining four convergent functions for $F(H)$ from three different gravitational theories, it is interesting to notice that, for the particular case when $\beta=1$, the four functions (\ref{GILA-convergent}),(\ref{beta-convergent}), (\ref{pure1-convergent}), and (\ref{pure2-convergent}), reduce to the same:

\begin{equation}\label{FdeH-beta1-unificada}
    F(H)=H^2+\lambda L^{2(p-1)}H^{2p}e^{\lambda(LH)^{2q}}- H^2 e^{-(\L H)^{2s}}.
\end{equation}

\section{Conclusions}

We explore a new approach to expressing the theory coined as Generalized Quasi-topological Gravity by considering an arbitrary coupling constant $\alpha_1$ on the Ricci curvature scalar in the action, i.e. $\alpha_1 R$ instead of the gravitational Einstein constant $1/(2\kappa) R$. We named that modification as  $\beta$-deformation theory. We showed that $\beta$-deformation is an extended theory of the proper GQTG. Later, we constructed the exponential convergent function for the modified Friedmann equations of both theories: $\beta$ theory and the proper GQTG. Next, we took $\alpha_1=0$, meaning that the linear Ricci curvature scalar is not considered in the GQGT action. In the end, we found two exponential convergent functions for this theory with no GR contribution that look similar to the cases GILA and $\beta$-deformation.

The fact that we found four convergent functions $F(H)$ from three different theories that are very similar between them makes us wonder how significant these differences in cosmology. Also, considering the apparent success of the cosmological analyses realized in \cite{Jaime:viability, Jaime:AUnified} for the GILA model, it is of interest to begin with the cosmological statistical exploration of the GILA, $\beta$-deformation theory and the $\beta=1$ \cite{Matias}.

Also, after observing that the cosmological scenario allows the theory to subtract the Ricci curvature scalar from the action and, despite that, the convergent function $F(H)$ maintains an $H^2$ contribution, makes us wonder what other astrophysical scenarios could be possible to be described with only the higher curvature order densities $\mathcal{R}^{(n)}$ playing a role. 



\section*{Acknowledments}

In some cases, calculations in this manuscript have been computed utilizing the Wolfram Mathematica packages xAct \cite{xact}. GA and LGJ acknowledge the financial support of SECIHTI-SNII. LGJ thanks the financial support of CONACYT-140630.

\appendix

 \bibliographystyle{elsarticle-num} 
 \bibliography{gila-biblio}





\end{document}